\documentclass{aa}
\usepackage{graphicx}
\begin{document}
\title{A new Wolf-Rayet star in Cygnus\thanks{Based on observations
collected at the German-Spanish Astronomical Centre, Calar Alto,
operated by the Max-Planck-Institut f\"ur Astronomie, Heidelberg, jointly 
with the Spanish National Commission for Astronomy.}
}
\author{Anna Pasquali
        \inst{1}
        \and
        Fernando Comer\'on
        \inst{2}
        \and
        Roland Gredel
        \inst{3}
        \and
    Jordi Torra
        \inst{4}
    \and
    Francesca Figueras
    \inst{4}}
\offprints{apasqual@eso.org}
\institute{ESO/ST-ECF,
             Karl-Schwarzschild-Strasse 2, 85748
             Garching bei M\"unchen, Germany\\
             \email apasqual@eso.org
           \and
           ESO,
             Karl-Schwarzschild-Strasse 2, 85748
             Garching bei M\"unchen, Germany\\
             \email fcomeron@eso.org
            \and
             Max-Planck Institut f\"ur Astronomie,
             K\"onigstuhl 17, D-69117 Heidelberg, Germany\\ 
             \email gredel@caha.es
             \and
             Departament d'Astronomia i Meteorologia, Universitat de
         Barcelona, Av. Diagonal 647, 08028 Barcelona, Spain \\
         \email jordi@am.ub.es, cesca@am.ub.es}
\date{Received; submitted}
\abstract{We report the discovery of a new Wolf-Rayet star in the direction of 
Cygnus. The star is strongly reddened but quite bright in the infrared, with 
$J = 9.22$, $H = 8.08$ and $K_S = 7.09$ (2MASS). On the basis of its H $+$ K 
spectrum, we have classified WR~142a a WC8 star. We have estimated its
properties using as a reference those of other WC8 stars in the solar
neighbourhood as well as those of WR~135, whose near-infrared spectrum
is remarkably similar. We thus obtain a foreground reddening of
$A_V \simeq 8.1$~mag, $M_J \simeq -4.3$, $\log (L/L_\odot) \sim 5.0-5.2$, 
$R = 0.8$~R$_\odot$, $T \simeq 125,000$~K, $M = 7.9-9.7$~M$_\odot$, 
${\dot M} = (1.4-2.3) \times 10^{-5}$~M$_\odot$~yr$^{-1}$. The
derived distance modulus, $DM = 11.2 \pm 0.7$~mag, places it in a
region occupied by several OB associations in the Cygnus arm, and
particularly in the outskirts of both Cygnus~OB2 and Cygnus~OB9. The position 
in the sky alone does not allow us to unambiguously assign the star to either
association, but based on the much richer massive star content of Cygnus~OB2 
membership in this latter association appears to be more likely.
\keywords{Stars: emission line -- Stars: fundamental parameters -- Stars:individual:Wolf-Rayet
-- Galaxy: open cluster and association}}
\titlerunning{A new Wolf-Rayet star in Cygnus}
\maketitle

\section{Introduction}

Wolf-Rayet (WR) stars, first discovered by Wolf and Rayet in 1867,
are characterised by an emission-line spectrum dominated by strong
He, N and C features. Their observed properties are believed to be
the result of stellar evolution at high initial masses, during
which strong mass loss in winds of a few 10$^{-5}$ M$_{\odot}$
yr$^{-1}$ at terminal velocities between 400 and 5000 km s$^{-1}$
(cf. van der Hucht 2001) ``peels'' the star until it comes to
expose at its surface the heavy elements synthesized in its core.
WR stars represent the last evolutionary stage of massive stars
before the final supernova explosion (Maeder \& Conti 1994, Langer
\& Woosley 1996, Langer \& Hegel 1999). The flux ratios among the
He, N, and C lines allows the classification of WR stars into
subclasses, WN (where He and N lines dominate), WC and WO (when
He, C and O features become stronger), and subtypes (WN3-9 and
WC4-9 as the ionization degree increases).

  WR stars significantly contribute to the energy budget and
chemical enrichment of their parent galaxy. Through their stellar
winds, they enrich the hosting environment with heavy elements and
release over time a kinetic energy comparable to that from
supernovae (Leitherer 1998). Given their high luminosities (few
10$^5$ L$_{\odot}$) individual WR stars are easily identified in
all the galaxies of the Local Group (Massey \& Johnson 1998),
while the short lifetimes of such massive stars (several Myr) make
them appear generally near their birthplaces, tracing star
formation in spiral and starburst galaxies. Indeed, their
unmistakable features, detected in the integrated spectra of
galaxies undergoing intense bursts of star formation (cf. Conti
2000), make them useful to study starburst phenomena and star
formation processes on cosmological scales.

The latest census of WR stars detected in the Milky Way lists 227
such objects (van der Hucht 2001), or about 10$\%$ of the
estimated Galactic WR population ($\simeq$ 1500 - 2500; Shara et
al. 1999). The majority of Galactic WR stars has escaped detection
using traditional techniques (such as visible spectroscopy or
narrow-band photometry with filters centered on the characteristic
WR signature at $\lambda \simeq$ 4670 \AA\/ and its adjacent
continuum, cf. Massey \& Conti 1983) which are naturally limited by
their unability to penetrate the large column densities of
absorbing dust along the lines of sight towards distant regions of
the galactic disk. Such limitation can be overcome by observing at
near-infrared wavelengths, where WR stars also display easily
identifiable spectral signatures and the extinction by dust is
only a fraction of that at shorter wavelengths.

In this paper we report the discovery of a new nearby WR star,
supporting the prediction by van der Hucht (2001) that many such
stars in the solar neighbourhood still await discovery at
near-infrared wavelengths. Indeed, the object reported here is
bright at near-infrared wavelengths but has escaped recognition so
far in the visible despite its proximity, due to the high
extinction in its direction. The discovery of yet another nearby
WR star deserves attention in several respects, as each WR star
has its own special features and represents a unique contribution
to our understanding of the late stages of massive stellar
evolution. Moreover, its location and relation to other young
stars in the region yields information on the star formation
history in the complex where it formed. The near-infrared
observations of the star, to which we will refer as WR~142a
following the nomenclature in van der Hucht's (2001) catalogue,
are described in Section 2. The results concerning its spectral
classification and intrinsic properties can be found in Section~3.
Finally, Section~4 discusses some implications of the WR
population in Cygnus.

\section{Observations}

The selection of WR~142a as a target for spectroscopic
observations was based on its 2MASS photometry, from which it was
apparent that this star is a bright, red object with colours
significantly deviating from those of normal stars reddened by
foreground dust. Infrared magnitudes listed by 2MASS are $J =
9.22$, $H = 8.08$, $K_S = 7.09$. Its location at $\alpha(2000) =
20^h 24^m 06.2^s$, $\delta(2000) =+41^\circ 25' 33''$ places it in
the general direction of Cygnus, in a region where several OB
associations overlap (see Section 3.4). The USNO catalog lists a
star of $B = 18.1$, $R = 14.7$ at the position of this object. The
only previous reference to it in the literature seems to be
Melikian \& Shevchenko (1990), who recognized it as a strong
emission-line star in their search for such objects in the
proximities of the cluster NGC~6910. However, its actual WR nature
has not been recognized until now.

We observed WR142a on June 24 and 25 2002, using the near-infrared
imager and spectrograph MAGIC mounted on the 1.23m telescope of
the German-Spanish Astronomical Center in Calar Alto (Spain). This
instrument uses a NICMOS3 $256 \times 256$ pixel$^2$ detector
yielding a scale of 1$''.$15 per pixel in imaging mode. $JHK_M$
imaging and $H$ and $K$ band spectroscopy were obtained and
reduced using the same instrumental setups and techniques
described in Comer\'on et al. (2002); the reader is thus referred
to that paper for details. Two spectra obtained on the night of 24
June 2002 at airmasses of 2.05 and 1.67 were individually reduced
to confirm the reality of individual features, especially in
regions of rapidly varying telluric absorption. The exposure time
of the combined spectrum is 12~min. As for the imaging, the
exposure times are 7.5~min in each of the $J$, $H$ and $K_M$
filters. The images were obtained under conditions suspected to be
non-photometric, and flux calibration using a separate image of a
standard star was thus not attempted.

\section{Results}

\subsection{IR spectroscopy and classification}

The low-resolution infrared spectrum of WR~142a ($R \simeq 240$) is
plotted in Figure~1. Emission lines have been identified by
cross-correlation with the The Atomic Line List v2.04
(http://www.pa.uky.edu/$\sim$peter/atomic) and the K-band spectral
atlas of Figer et al. (1997). The most prominent features are due
to HeI and HeII, CIII and CIV. Their equivalent widths are listed
in Table~1.

\begin{figure}
   \centering
   \includegraphics[width=8.5cm]{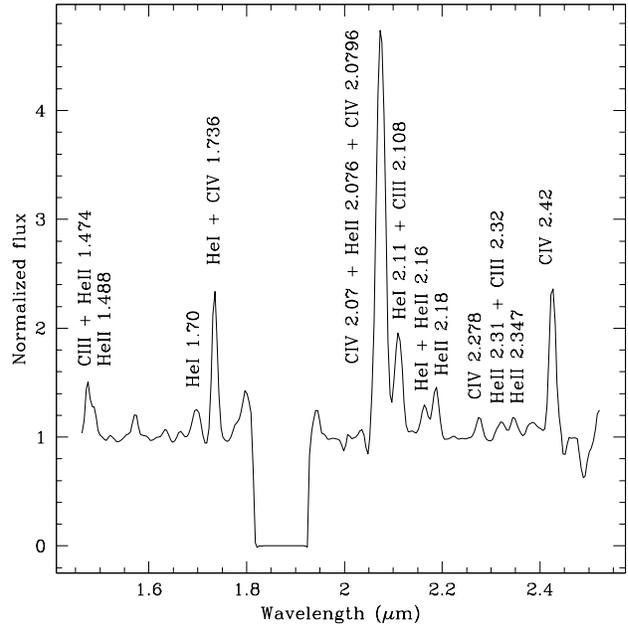}
   \caption{The observed H-to-K spectrum of WR~142a. The
    spectral resolution is R $\simeq$ 240. The drop between 1.8 and
    1.9 $\mu$m is due to the Earth atmosphere transmission.}
\end{figure}
\par\noindent

\begin{table}
\caption[]{Line equivalent widths in the IR spectrum of WR~142a}
\begin{tabular}{l c}
\hline
Line ($\mu$m)    & EqW (\AA) \\
\hline
\noalign{\smallskip}
CIII $+$ HeII 1.474 & 50 \\
HeII 1.488          & 30 \\
HeI 1.70            & 60 \\
HeI + CIV 1.736     & 200 \\
CIV 2.07 + HeII 2.076 + CIV 2.08 & 780 \\
HeI 2.11 + CIII 2.108 & 160 \\
HeI + HeII 2.16 & 40 \\
HeII 2.18 & 70 \\
CIV 2.278 & 30 \\
HeII 2.31 + CIII 2.32 & 30 \\
HeII 2.347 & 30 \\
CIV 2.42   & 200 \\
\hline
\end{tabular}
\end{table}

We have cross-checked the spectrum of WR~142a with the K-band
spectral atlas of Galactic WR stars by Figer et al. (1997) in
order to assign WR~142a a WR subtype. The line complex between 2
and 2.2 $\mu$m has been particularly useful in the spectral
classification, since it is a specific signature (in line
wavelength and flux ratio) of WC8 stars (cf. Figure 10 in Figer et
al. 1997). Although the dispersion of Figer et al.'s (1997) data
is a factor of two higher (R $\simeq$ 525 against 240 of this
work), the comparison indicates that the spectrum of WR~142a is
almost identical to that of WR~135, a WC8 star in the Galactic OB
association Cygnus OB3. And similarly to WR~135, a weak absorption
can be detected blueward of the CIV + HeII emission line at
$\lambda$ = 2.075 $\mu$m suggesting a P Cygni profile for this
line blend (or for the HeII 2.06 $\mu$m feature buried in this
line blend). Therefore, we classify WR~142a as a WC8 star.

\subsection{Colors, distance, reddening and variability}

As a first step towards the estimate of the physical properties
of WR~142a, we have compared its near-infrared photometry to that of
six other nearby WC8 stars (WR~53, WR~60, WR~77, WR~101, WR~117,
and WR~135) selected from Williams et al. (1987). Stars with
detected companions were excluded. WR~101 and WR~135 have $JHK$
photometry also from 2MASS, and WR~113 from Pitault et al. (1983);
in all these cases, all published magnitudes agree to better than
0.1~mag. Extinctions in the visible, $A_V$, are given for the six
reference objects by van der Hucht (2001) based on their visible
photometry. A seventh WC8 star, WR~102e, also has available 2MASS
photometry. However, the fact that this star is located in the
galactic center raises some concerns about its validity as a
reference for comparison to the WR population in the solar
neighbourhood. Besides the published $JHK$ photometry we have also
used the extinction given in van der Hucht catalogue, from which
we derive intrinsic colours $(J-H)_0$, $(H-K)_0$ using the
extinction law of Rieke \& Lebofsky (1985).

Assuming that the intrinsic colours of WR~142a are the same as
the average of those of the reference stars, we obtain $(J-H)_0 =
0.28 \pm 0.25$. The uncertainty includes both the intrinsic
scatter in colours among WC8 stars and possible errors in
determining the extinction, which is generally done using $BV$
colours. The derived extinction for WR~142a is thus $A_V = 8.1 \pm
2.3$, or $A_J = 2.3 \pm 0.6$. Using the $K$ band to estimate the
extinction introduces larger uncertainties due to the larger
scatter in derived intrinsic $(H-K)_0$ colours among the reference
objects, and we have preferred to exclude it for this purpose.
A more accurate estimate is possible by using the measured $B = 18.1$
in the USNO catalogue, keeping in mind that a possible error
may be introduced by variability in the non-simultaneous $B$ and $JHK$
observations. In that case, we have calculated $(B-V)_0 = -0.14$ using
the observed $(B-V)$ of the reference stars and dereddening them with
the $A_V$ listed by van der Hucht (2001), using $A_V/(A_B-A_V) = 3.1$,
thus obtaining $(B-J)_0 = 0.42$. The observed $(B-J) = 8.9$ thus implies
$A_V = 8.1$, practically identical to the value derived from $J-H$ but 
based on a much longer wavelength baseline. The uncertainty, mostly due
to the scatter in $(B-J)_0$, is accordingly smaller and estimated to
be 0.4~mag. We thus adopt $A_V = 8.1 \pm 0.4$, or $A_J = 2.3 \pm 0.1$, 
as the extinction towards WR~142a.

A similar procedure applied to the $(V-J)$ photometry of the 
reference stars yields an estimate of $(V-J)_0 \simeq 0.56 \pm 0.24$ 
for WR~142a, again with the caveat of the non-simultaneity of the 
visual and infrared observations although the relatively 
small scatter that we find in $(V-J)_0$ suggests that this may not 
be an important factor. Proceeding with the $(V-J)_0$ given above 
and assuming for the visual absolute magnitude of WR~142a the mean 
value derived by van der Hucht (2001) for WC8 stars, $M_V = -3.74$, we 
obtain $M_J = -4.30$. In this case, to the already noted uncertainty 
in $(V-J)_0$ we must add the scatter in $M_V$ found for these stars,
which van der Hucht evaluates at 0.5~mag. Using the estimates of
the extinction and the absolute magnitude we obtain a distance
modulus of $DM = 11.2 \pm 0.7$, corresponding to a distance of
$1.8^{+0.6}_{-0.5}$~kpc.

  An alternative possibility consists of using the close similarity
between the spectra of WR~142a and WR~135 to assume that the
intrinsic properties of both are identical, including intrinsic
colours and absolute magnitudes. WR~135 is the bluest star in the
reference sample both in $(J-H)_0$ and $(V-J)_0$. The brighter
$M_V$ listed by van der Hucht (2001) is balanced by the bluer
colours to give $M_J = -4.33$, practically identical to the value
used in the above estimate. However, fitting the colours requires a
higher extinction ($A_V = 11.0$ or $A_J = 3.1$) and the distance
modulus thus reduces to 10.2~mag, corresponding to a distance of
1.2~kpc, in the lower end of the range given above.

  Despite of the fact that our observations were probably performed
under non-photometric conditions  as noted in Section 2, it is
still possible to check for possible variability by comparing
WR~142a to other 2MASS stars in the field. Such comparison
suggests that WR~142a was slightly fainter when we observed it
than when it was observed by 2MASS, with $\Delta J = 0.08$,
$\Delta H = 0.07$, $\Delta K = 0.25$. Only the significance of the
$K$ variability is more than marginal, lying within the amplitude
range of the long-term variability of Wolf-Rayet stars reported by
Hackwell et al. (1979) and well below the rather extreme case
described by Danks et al. (1983).

\subsection{Properties and environment}

The physical properties of WR 142a (mass $M$, radius $R$,
temperature $T$) may be estimated with simple
relations derived by Schaerer \& Maeder (1992) from their models
of WNE/WC stars, with the caveat that these models
are quite sensitive to the assumed morphology of the stellar wind,
i.e. clumpiness, which may change the bolometric luminosity of
WRs. The luminosity can be estimated by
adopting for WR~142a $M_V = -3.74$ as determined by van der Hucht
(2001) for WC8 stars, and the average bolometric correction $BC_V
= -4.12$ found by Nugis \& Lamers (2000) for the WC8 subgroup (taking
into account their equation (6) to transform from narrow- to broad-band
$V$ magnitude). Since $\log(L/L_\odot) =
-0.4 \times (M_V + BC_V - M^{\rm BOL}_{\odot})$, the total
luminosity of WC8 stars turns out to be $\log(L/L_\odot) = 5.04$.
If the individual luminosity of WR~135 is adopted instead as more
appropriate for WR~142a, we obtain $\log(L/L_{\odot}) = 5.24$.

  Using Schaerer and Maeder's mass-luminosity relation,

$$\log(L/L_{\odot})=3.032+2.695
\log(M/M_{\odot})-0.461(\log(M/M_{\odot}))^2$$

\noindent we obtain a mass between 7.9 and 9.7~M$_\odot$ (assuming
respectively $\log L/L_{\odot} = 5.04$ and 5.24 as described
above). Likewise, the stellar radius $R$ and temperature $T$ are
given approximated by

$$\log(R/R_{\odot}) = -1.845 + 0.338 \times \log(L/L_{\odot})$$

\noindent and

$$\log(T) = 4.684 + 0.0809 \times \log(L/L_{\odot})$$

\noindent from where we obtain $R \simeq 0.8$~R$_\odot$, $T \simeq
125000$~K, with little dependence on the choice between 5.04 or
5.24 for $\log L/L_{\odot}$. Finally, adopting Nugis \& Lamers'
equation (24), which gives the stellar mass loss as a function of
the stellar present mass, we derive for WR142a a mass-loss rate of
$1.4 \times 10^{-5}$~M$_\odot$~yr$^{-1}$ or $2.3 \times
10^{-5}$~M$_\odot$~yr$^{-1}$, depending on which one of the masses
given above is adopted. It must be stressed that the values above
are a simple estimate of the stellar properties of WR142a based on
average properties of WC8 stars, and an accurate model atmosphere
fitting of its IR spectrum is indeed needed.

  Although the winds of late-type WC stars are generally a source of
warm dust (Smith \& Houck 2001, Chiar et al. 2001), no IRAS
mid-infrared point source is detected near the position of
WR~142a. This is somewhat unexpected, as the sensitivity of IRAS
should enable detection of a typical late WC star within 7~kpc
from the Sun (Cohen 1995). Mid-infrared observations provide no
evidence either for any extended structure at larger scales around
WR~142a, as is sometimes found the neighbourhood of other WR stars
(e.g. Miller \& Chu 1993, Saken et al. 1995, Pineault \& Terebey
1997, Gervais \& St-Louis 1999, and references therein). This does
not completely rule out the possible existence of an extended
nebula, as the wide range of ring nebula sizes around WR stars
(Miller \& Chu 1993) and the intricate structure of thermal
infrared emission in the general direction of Cygnus would
naturally make the detection of such a nebula elusive.
Near-infrared images of the immediate surroundings of WR~142a
(Figure~2) also show it to be rather inconspicuous, with no
obvious nebulosity associated to it within the $5'$ field of view
of our images.

\begin{figure}
   \centering
   \caption{A view of the field around WR~142a (the bright star at
   the center of the image) showing no evidence for extended
   emission. The printed version of this paper shows the $K$-band
   image, while a $JHK$ colour composite is presented in the
   electronic version.}
\end{figure}

\subsection{Membership}

  At the estimated distance of $\simeq 1.8$~kpc, WR~142a lies in
the extended complex of massive star forming regions composed by
several OB associations and part of the molecular complex Cygnus~X
(e.g. Odenwald \& Schwartz 1989). Its location with respect to the
three most nearby associations and the other WR stars identified
within $5^\circ$ of it are shown in Figure~3.

\begin{figure}
   \centering
   \includegraphics[width=8.5cm]{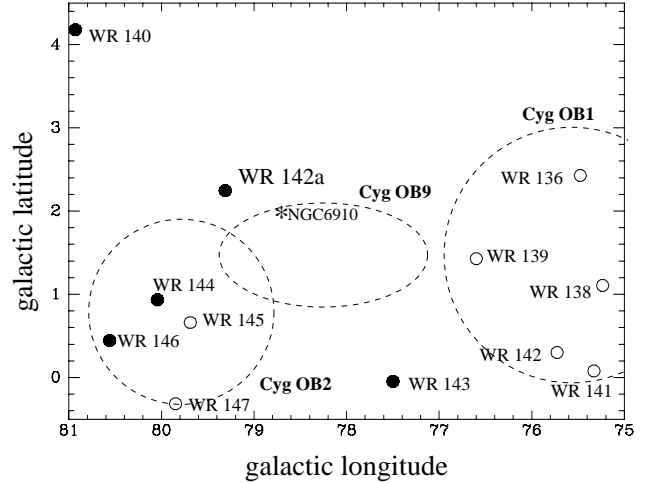}
   \caption{Location of WR~142a with respect to the three OB
   associations in its neighbourhood. Also shown are the positions
   of other WR stars, with full circles representing WC stars and
   empty circles WN stars. The boundaries of the associations are
   a rough approximation to the distribution of their members as
   plotted in Garmany \& Stencel (1992).}
\end{figure}

  The overlap among the associations and the difficulty in precisely
establishing their boundaries (or even in establishing them as
different entities) makes it difficult to unambiguously assign
WR~142a to one of them. Adopting the distribution of members given
by Garmany \& Stencel (1992), WR~142a lies near the boundaries of both
Cygnus~OB2 and Cygnus~OB9. Association to the latter may be suggested
in view of its proximity (less than $1^\circ$) to the open cluster
NGC~6910. The average extinction towards Cygnus~OB9, where members
obscured by $A_V > 6$ are listed by Garmany \& Stencel (1992),
does not conflict with this assignment. It may be interesting to
note that WR~142a appears in the sky about $10'$ West of the
compact HII region DWB87 (Dickel et al. 1969) and may be obscured
by the molecular cloud surrounding it. No detailed studies of this
nebula seem to exist in the literature, preventing us from
ascertaining whether a physical connection between WR~142a and
DWB87 may exist. On the other hand, it is also possible that 
WR~142a is actually a member of Cygnus~OB2. Cygnus~OB2
is a compact OB association (see Kn\"odlseder 2000, Comer\'on et
al. 2002, and references therein) rich in very massive members
which are on the average more reddened than those of Cygnus~OB9.
Cygnus~OB2 has four other WR stars, two of which also belong to the
WC class, indicating progenitors with initial masses above
60~M$_\odot$ (Crowther et al. 1995a), and in this respect the
higher content of very massive stars of Cygnus~OB2 favors the
hypothesis that WR~142a is a member of this association lying in
its outskirts. Given the considerable uncertainties affecting the
definition of the membership, the boundaries, and the distances of
each of these associations (see Comer\'on et al. 1998 for an
additional discussion), the assignment of WR~142a to either
Cygnus~OB2 or Cygnus~OB9 must remain as only tentative for the time
being.

\section{Discussion and conclusions}

  Our near-infrared observations of a source in Cygnus with infrared
colours deviating from those of normal, reddened stars has brought
up the serendipitous discovery of another WR star in this complex
of rich OB associations. Table~2 summarizes the parameters of this
star, WR~142a.

\begin{table}
\caption[]{Summary of properties of WR~142a}
\begin{tabular}{lc}
\hline \noalign{\smallskip}
$\alpha(2000)$ & $20^h 24^m 06.2^s$ \\
$\delta(2000)$ & $+41^\circ 25' 33''$ \\
\noalign{\smallskip}
$J$~(2MASS) & 9.22 \\
$H$~(2MASS) & 8.08 \\
$K$~(2MASS) & 7.09 \\
\noalign{\smallskip}
$J$~(June 2002) & 9.31 \\
$H$~(June 2002) & 8.15 \\
$K$~(June 2002) & 7.34 \\
\noalign{\smallskip}
spectral type & WC8 \\
\noalign{\smallskip}
$r$~(kpc) & $1.8^{+0.6}_{-0.5}$ \\
$A_V$~(mag) & $8.1 \pm 0.4$ \\
\noalign{\smallskip}
$\log (L/L_\odot)$ & 5.04$^1$ \\
$M$~(M$_\odot$) & 7.9$^1$ \\
$R$~(R$_\odot$) & 0.8$^1$ \\
$T$~(K) & 125,000$^1$ \\
\hline
\end{tabular}
\\
$^1$: Quantities derived from average properties of WC8
stars\\
\end{table}

  The detection of WR~142a adds one more example of a late evolved star
to the massive star content of this region. It is a remarkable
fact that Cygnus~OB2 contains examples of all the stages in the
evolutionary path of the most massive stars,
O main sequence $\rightarrow$ Of/WN $\rightarrow$ LBV
$\rightarrow$ WN $\rightarrow$ WC for initial masses about 
60 M$_{\odot}$ (Langer et al. 1994). The addition of one more
member of the rare class of stars with such high initial masses is
thus a valuable one to a region that can be considered as the best
avaliable laboratory for the observational study of stellar
evolution at the upper end of the main sequence.

  The abundant massive stars content of the complex Cygnus~OB2/OB9,
where over 100 O-type stars and descendants have been identified
(Garmany \& Stencel 1992, Massey et al. 1995, Kn\"odlseder 2000,
Comer\'on et al. 2002), leads to the expectation that this complex
may be at present or in the near future the scenario of intense
supernova activity. Taking 7.1~Myr as the expected lifetime of a
late O-type star (Meynet et al. 1994, Schaerer \& de Koter 1997), 
the content of Cygnus~OB2/OB9 would lead to the expectation of a rate 
of about 1 SN every less than 70,000 yr. The observability of the
radio-continuum signature of individual supernova remnants is
limited in time and highly dependent on selection effects that are
difficult to model (Green 1991), but the inferred interval between
supernovae in Cygnus is rather short as compared to the estimated 
age of many supernova remnants (Matthews et al. 1998). Thus, the 
fact that none has been observed
within 2 degrees from the centre of Cygnus~OB2 (cf. Green 2001)
suggests that no evolved massive star has gone through the
supernova phase yet. This poses some constraints on the age of
Cygnus OB2. According to the evolutionary track for an initial
mass of 60 M$_{\odot}$ computed by Langer et al. (1994), a star
enters the WC phase when its mass has decreased to 15.3
M$_{\odot}$. Only when its present mass is as small as 3.92
M$_{\odot}$, the star can be considered a probable candidate for a
supernova explosion. This happens after 4.065 $\times$ 10$^6$ yrs
from the main-sequence phase. Moreover, according to Langer et al.
(1994), WC stars pop up around 3 million years. Hence, we may argue 
that Cygnus OB2 is as young as 3 - 4 million years, basically in  
agreement with Massey et al.
(2001) who derived, from photometry of OB main-sequence stars, a
mean age of 2.6 $\times$ 10$^6$ yrs.

An inspection of van der Hucht's (2001) catalogue reveals that
Cygnus~OB2 is the richest association in number of WC stars compared
to WN stars, and the discovery of the new WC star reported here
underlines this difference with respect to other associations.
This could be interpreted as the result of a shallower IMF combined
with the young age of the association. Indeed, an IMF slope of -0.9
has been measured for Cygnus~OB2 by Massey (1998) and the cluster
turn-off mass has been determined of about 120 M$_{\odot}$ by
Massey et al. (2001).

  The discovery of a new Wolf-Rayet at a distance less
than 2~kpc from the Sun reported here dramatically illustrates the
incompleteness that affects the census of even the most massive
stars in our galactic neighbourhood, while stressing the need for
dedicated infrared surveys to improve the sample. Such efforts
will be certainly worthwhile, given the critical dependence that
our understanding of the star formation history and stellar mass
function in OB associations has on working on complete samples,
especially when the most massive and rarest members are concerned.

\begin{acknowledgements}
We would like to thank an anonymous referee whose comments
improved this manuscript. Also,
it is a pleasure to thank the staff of the Calar Alto observatory,
especially Mr. Santos Pedraz, for their assistance during the
observing run in which the observations presented here were
obtained. We also appreciate the help of Ms. Uta Grothkopf in
making available to us the original article of Melikian and
Shevchenko, and of Ms. Petia Andreeva in translating it from
Russian. This paper made use of data obtained as part of the Two Micron 
All Sky Survey (2MASS), a joint project of the University of Massachusetts
and the Infrared Processing and Analysis Center/California Institute of 
Technology, funded by the National Aeronautics and Space Administration 
and the National Science Foundation; and of the SIMBAD database, operated 
at CDS, Strasbourg, France.
\end{acknowledgements}

\end{document}